\documentclass[twocolumn,showpacs,aps,epsfig,nofootinbib]{revtex4}

\usepackage{graphicx}
\usepackage{amsfonts}
\usepackage{epstopdf}
\usepackage{latexsym}
\usepackage{amssymb}
\usepackage{amssymb}

\usepackage[center]{subfigure}

\begin{document}

 \newcommand{\bq}{\begin{equation}}
 \newcommand{\eq}{\end{equation}}
 \newcommand{\bqn}{\begin{eqnarray}}
 \newcommand{\eqn}{\end{eqnarray}}
 \newcommand{\nb}{\nonumber}
 \newcommand{\lb}{\label}
\newcommand{\PRL}{Phys. Rev. Lett.}
\newcommand{\PL}{Phys. Lett.}
\newcommand{\PR}{Phys. Rev.}
\newcommand{\CQG}{Class. Quantum Grav.}

\title{Static post-Newtonian limits in non-projectable   Ho\v{r}ava-Lifshitz gravity with an extra U(1) symmetry}

\author{Kai Lin ${}^{a,b}$}
\email{lk314159@hotmail.com}

 \author{Anzhong Wang ${}^{a, c}$} 
\email{anzhong_wang@baylor.edu}

\affiliation{ ${}^{a}$ Institute  for Advanced Physics $\&$ Mathematics,   Zhejiang University of
Technology, Hangzhou 310032,  China \\
${}^{b}$  Instituto de F\'isica, Universidade de S\~ao Paulo, CP 66318, 05315-970, S\~ao Paulo, Brazil \\
${}^{c}$ GCAP-CASPER, Physics Department, Baylor
University, Waco, TX 76798-7316, USA }

\date{\today}

\begin{abstract}

In this paper, we study static post-Newtonian limits in non-projectable   Ho\v{r}ava-Lifshitz gravity with an extra U(1)
symmetry. After obtaining  all static spherical  solutions  in the infrared, we apply them to the solar system tests, and
obtain  the Eddington-Robertson-Schiff parameters in terms of the coupling constants of the theory. These parameters
are well consistent with observations for the physically viable coupling constants. In contrast to the projectable case,
this consistence is achieved without taking the gauge field and Newtonian prepotential as part of the metric.

\end{abstract}


\maketitle

\section{Introduction}
\renewcommand{\theequation}{1.\arabic{equation}} \setcounter{equation}{0}

To quantize gravity in the framework of quantum field theory, recently Ho\v{r}ava proposed the Ho\v{r}ava-Lifshitz (HL) theory
of gravity \cite{Horava}, in which the Arnowitt-Deser-Misner  (ADM) variables   \cite{ADM}
are taken as  the fundamental quantities to describe  gravity.
By construction, the theory is power-counting renormalizable, which is realized by including
high-order spatial derivative operators [up to six in (3+1)-dimensional spacetimes].
The exclusion of high-order time derivative operators, on the other hand, ensures
that the theory is unitary, a problem that has been faced in high-order  derivative theories of gravity for a long time \cite{Stelle}.
Clearly, this inevitably breaks the general  diffeomorphisms,
\bq
\lb{A.0a}
\delta{x}^{\mu} = -\zeta^{\mu}(t, x) , (\mu = 0, 1,  2,  3).
\eq
Although  such a
breaking  in the gravitational sector is much less restricted by experiments/observations than that in the matter sector  \cite{LZbreaking,Pola},
it is still a challenging question how to prevent the propagation of the Lorentz violations into the Standard Model of particle physics \cite{PS}.
Ho\v{r}ava  assumed that such a breaking only happens in the ultraviolet (UV) and down to the foliation-preserving diffeomorphism,
\bq
\lb{A.0b}
 \delta{t} = - f(t), \;\;\; \delta{x}^i = - \zeta^i(t, x), (i = 1, 2, 3),
\eq
often denoted by ${\mbox{Diff}}(M, \; {\cal{F}})$. In the infrared (IR), the low derivative operators take over, presumably providing a
healthy low energy limit.

The breaking of the  general  diffeomorphisms immediately results in the appearance of spin-0 gravitons in the theory, in addition to
the spin-2 ones, found in general relativity (GR). This is potentially dangerous, and leads to several problems, including instability,  strong coupling and
different speeds of (massless) particles \cite{reviews}. To resolve these problems, various models have been proposed \cite{reviews}, including the
healthy extension of the non-projectable HL theory \cite{BPS}. In the healthy extension, the instability problem was fixed by the inclusion
of the term $\beta_0 a_i a^i$ in the gravitational action,  where $\beta_0$ must be in the range $ 0 < \beta_0 < 2$, and $a_i \equiv N_{,i}/N$ with
$N$ being the lapse function in the ADM decompositions \cite{ADM}.
The strong coupling problem is resolved by introducing a new energy scale $M_{*}$, so that $M_{*} \leq \Lambda_{\omega} \equiv \sqrt{\beta_0} M_{pl}$, where
$M_{*}$ denotes the suppression energy of the high-order spatial operators, $\Lambda_{\omega}$ the would-be strong coupling energy scale,
and $M_{pl}$ the Planck mass. Clearly, in order for this mechanism
to work, one must assume that $\beta_0 \not= 0$. Observational constraints requires $10^{10}\; {\mbox{GeV}}\; \leq M_{*} \leq 10^{15}\; {\mbox{GeV}}$
\cite{BPS}. The low bound was obtained by assuming that $M_{*}$ also sets the suppression energy scale in matter fields, while the up bound
was obtained from the preferred frame effects \cite{Will,FJ}. In addition, to avoid the Cherenkov radiation, one must require that the speed of the spin-0
gravitons be superluminal  \cite{EMS}  \footnote{This can be easily seen using the equivalence between the Einstein-aether theory and the nonprojectable HL theory
in the IR limit \cite{BPS,Jacobson}.}.

A more dramatical  modification was proposed recently by  Ho\v{r}ava and Melby-Thompson (HMT) \cite{HMT}, in which  an extra local U(1)
symmetry was introduced, so that the symmetry of the theory was enlarged to,
\bq
\lb{symmetry}
 U(1) \ltimes {\mbox{Diff}}(M, \; {\cal{F}}).
\eq
Such an extra symmetry is realized by introducing a gauge field $A$ and a Newtonian prepotential $\varphi$.
Because of this extra symmetry,  the spin-0 gravitons are eliminated \cite{HMT,WWa}. As a result, all the problems related to them, such as the
instability, strong coupling and different speeds in the gravitational sector, as mentioned above, are automatically resolved. This was initially
done with the projectability condition $N = N(t)$ and $\lambda = 1$ \cite{HMT}, where $\lambda$ characterizes the IR deviation of the theory from
GR. It was soon  generalized to the case with any value of $\lambda$ \cite{daSilva},  in  which   the spin-0 gravitons are still eliminated
\cite{daSilva,HW}. Although the strong coupling problem in the gravitational sector disappears, it still exists in the matter sector \cite{HW},
but can be resolved also by  introducing a new energy scale $M_{*}$, so that  $M_{*} \leq |\lambda -1|^{1/4} M_{pl}$ \cite{LWWZ}. Cosmological
applications of this model were considered in \cite{HWW,HW2}, and found that it is consistent with current observations \cite{Cosmo}. On the other hand,
the studies of solar system tests recently showed that the theory is consistent with observations {\em if and only if}
the gauge field  $A$ and the Newtonian prepotential $\varphi$ are part of the metric \cite{LMW}  \footnote{When $\lambda = 1$,
the theory is consistent with the solar system tests even without the gauge field  and the Newtonian prepotential being part of the metric \cite{AP,GSW},
but the corresponding cosmology is quite different from the standard one \cite{HW}. Other considerations of this model can be found in \cite{BLW,daSilva2}.}.

A non-trivial generalization of the enlarged symmetry (\ref{symmetry}) to the nonprojectable case $N = N(t, x)$ was recently worked out in \cite{ZWWS,ZSWW},
and showed that the only degree of freedom of the model in the gravitational sector is the spin-2 massless gravitons, the same as that in GR. Because
of the elimination of the spin-0 gravitons, the physically viable region of the coupling constants is considerably enlarged, in comparison with the healthy
extension \cite{BPS}, where   the extra U(1) symmetry is absent.   In particular, the requirement $\beta_0 \not= 0$ now is dropped out. Furthermore, the
number of independent coupling constants is also dramatically reduced, from about 100 down to 15.  The consistence of the model with cosmology was
showed recently in \cite{ZSWW,ZHW,WWZZ}, and various remarkable features were found.

In this paper, we  study the consistence of the  model proposed in \cite{ZWWS,ZSWW} with the solar system tests. In particular, we  investigate its
 static post-Newtonian limits. The paper is organize  as follows: In the next section (Sect. II), we give a
brief introduction to the model, and in Sec. III we find the spherical static solutions in the IR limit, from which we can see that the Birkhoff theorem is not
applicable to the current theory, because of the breaking of the Lorentz symmetry. As a results, there are several classes of static asymptotically
flat vacuum solutions.
Then, in Sec. IV, we study the post-Newtonian  limit for each class of these solutions, and find the Eddington-Robertson-Schiff parameters, $\gamma$ and $\beta$,
in terms of the coupling constants of the theory,  where $\gamma$ is related to the amount of  spatial curvature generated by the spherical source, and $\beta$ to the
degree of non-linearity in the gravitational field.  By doing so, we show that the observational constraints on $\gamma$ and $\beta$ can be easily satisfied
within the physically viable region of the phase space of the  coupling constants of the theory. This is true without taking the gauge field $A$ and Newtonian
prepotential $\varphi$ as part of the metric, in contrast to the   projectable case \cite{LMW}.
In Sec. V, we summarize our main results and present some discussions and remarks.

\section {Non-Projectable HL Theory with U(1) Symmetry}
\renewcommand{\theequation}{2.\arabic{equation}} \setcounter{equation}{0}

In this section, we shall give a brief review of the  non-projectable HL theory with the enlarged symmetry (\ref{symmetry}) \cite{ZWWS,ZSWW}.
The fundamental variables of the theory are $\left(N, \; N^i, \; g_{ij}, \; A, \; \varphi\right)$,  where $N^i$ and $g_{ij}$ are, respectively, the  shift vector, and 3-metric
of the leaves $t = $ Constant  in the ADM decompositions \cite{ADM}.  Under the local $U(1)$ symmetry,
the   fields   transform as
\bqn
\lb{A.0d}
\delta_{\alpha}A &=&\dot{\alpha} - N^{i}\nabla_{i}\alpha,\;\;\;
\delta_{\alpha}\varphi = - \alpha,\nb\\
\delta_{\alpha}N_{i} &=& N\nabla_{i}\alpha,\;\;\;
\delta_{\alpha}g_{ij} = 0 = \delta_{\alpha}{N},
\eqn
where $\alpha$ is   the generator  of the local $U(1)$ gauge symmetry, $\dot{\alpha} \equiv \partial\alpha/\partial t$,
and $\nabla_i$ the covariant derivative with respect to the 3-metric $g_{ij}$.
Under the Diff($M, \; {\cal{F}}$), they transform as,
\bqn
\lb{A.0e}
\delta{N} &=& \zeta^{k}\nabla_{k}N + \dot{N}f + N\dot{f},\nb\\
\delta{N}_{i} &=& N_{k}\nabla_{i}\zeta^{k} + \zeta^{k}\nabla_{k}N_{i}  + g_{ik}\dot{\zeta}^{k}
+ \dot{N}_{i}f + N_{i}\dot{f}, \nb\\
\delta{g}_{ij} &=& \nabla_{i}\zeta_{j} + \nabla_{j}\zeta_{i} + f\dot{g}_{ij}, \nb\\
\delta{A} &=& \zeta^{i}\nabla_{i}A + \dot{f}A  + f\dot{A},\nb\\
\delta \varphi &=&  f \dot{\varphi} + \zeta^{i}\nabla_{i}\varphi.
\eqn

The general action reads \cite{ZWWS,ZSWW},
\bqn
\label{action}
S&=&\zeta^2 \int dtd^3x \sqrt{g} N \Big({\cal{L}}_{K}-{\cal{L}}_{V} + {\cal{L}}_{A}
+ {\cal{L}}_{\varphi}\nb\\
&& ~~~~~~~~~~~~~~~~~~~~~~~~~  + {\zeta^{-2}} {\cal{L}}_M\Big),
\eqn
where $\zeta^2 = 1/(16\pi G)$ with $G$ being the Newtonian constant, ${\cal{L}}_M$ describes matter fields, and
\bqn
\lb{2.2}
{\cal{L}}_{K} &=& K_{ij}K^{ij} - \lambda K^2,\nb\\
{\cal{L}}_{V} &=&  {\cal{L}}_{V}^{R} + {\cal{L}}_{V}^{a},\nb\\
{\cal{L}}_{A} &=& \frac{A}{N}\left(2\Lambda_g - R\right),\nb\\
{\cal{L}}_{\varphi} &=&  \varphi{\cal{G}}^{ij}\big(2K_{ij}+\nabla_i\nabla_j\varphi+a_i\nabla_j\varphi\big)\nb\\
& & +(1-\lambda)\Big[\big(\Delta\varphi+a_i\nabla^i\varphi\big)^2\nb\\
&&  ~~~~~~~~~~~~ ~ +2\big(\Delta\varphi+a_i\nabla^i\varphi\big)K\Big]\nb\\
& & +\frac{1}{3}\hat{\cal G}^{ijlk}\Big[4\left(\nabla_{i}\nabla_{j}\varphi\right) a_{(k}\nabla_{l)}\varphi \nb\\
&& ~~~~~~~~~~~ ~ + 5 \left(a_{(i}\nabla_{j)}\varphi\right) a_{(k}\nabla_{l)}\varphi\nb\\
&&
+ 2 \left(\nabla_{(i}\varphi\right)a_{j)(k}\nabla_{l)}\varphi
+ 6K_{ij} a_{(l}\nabla_{k)}\varphi\Big],
\eqn
with $\Delta \equiv \nabla^2$, and
\bqn
\lb{2.2b}
K_{ij} &=& \frac{1}{2N}\left(- \dot{g}_{ij} + \nabla_{i}N_{j} +  \nabla_{j}N_{i}\right),\nb\\
a_{i} &=& \frac{N_{,i}}{N},\;\;\; a_{ij} = \nabla_{j} a_{i},\nb\\
\hat{\cal G}^{ijlk} &=& g^{il}g^{jk} - g^{ij}g^{kl}, \nb\\
{\cal{G}}_{ij} &=& R_{ij} - \frac{1}{2}g_{ij} R + \Lambda_g g_{ij},\nb\\
{\cal{L}}_V^R&=&\gamma_0 \zeta^2 + \gamma_1 R+\frac{\gamma_2 R^2+\gamma_3 R_{ij}R^{ij}}{\zeta^2}+\frac{\gamma_5}{\zeta^4} C_{ij}C^{ij},\nb\\
{\cal{L}}_{V}^{a} &=&    -  \beta_0  a_{i}a^{i}
  + \frac{1}{\zeta^{2}}\Bigg[\beta_{1} \left(a_{i}a^{i}\right)^{2} + \beta_{2} \left(a^{i}_{\;\;i}\right)^{2} \nb\\
& & + \beta_{3} \left(a_{i}a^{i}\right)a^{j}_{\;\;j} + \beta_{4} a^{ij}a_{ij} + \beta_{5}
\left(a_{i}a^{i}\right)R \nb\\
& &   + \beta_{6} a_{i}a_{j}R^{ij} + \beta_{7} Ra^{i}_{\;\;i}\Bigg]  +  \frac{1}{\zeta^{4}}\  \beta_{8} \left(\Delta{a^{i}}\right)^{2}.
\eqn
Here $R_{ij},\ R$ are, respectively the Ricci tensor and scalar of $g_{ij}$, and $C_{ij}$ denotes the
 Cotton tensor, defined by
\bq
\lb{2.2c}
C^{ij} =  \frac{e^{ijk}}{\sqrt{g}} \nabla_{k}\Big(R^{j}_{l} - \frac{1}{4}R\delta^{j}_{l}\Big),
\eq
with  $e^{123} = 1$, etc. $\lambda, \gamma_n, \beta_{s} $ and $ \Lambda_g$   are the coupling constants of the theory. In terms of $R_{ij}$, we have
\cite{ZSWW},
\bqn
\lb{2.2da}
C_{ij}C^{ij}
&=& \frac{1}{2}R^{3} - \frac{5}{2}RR_{ij}R^{ij} + 3 R^{i}_{j}R^{j}_{k}R^{k}_{i}  +\frac{3}{8}R\Delta R\nb\\
& &  +
\left(\nabla_{i}R_{jk}\right) \left(\nabla^{i}R^{jk}\right) +   \nabla_{k} G^{k},
\eqn
where
\lb{2.2e}
\bqn
G^{k}=\frac{1}{2} R^{jk} \nabla_j R - R_{ij} \nabla^j R^{ik}-\frac{3}{8}R\nabla^k R.
\eqn

The infrared limit requires
that
\bq
\lb{2.2d}
\Lambda = \frac{1}{2}\gamma_0 \zeta^2, \;\;\; \gamma_1  = -1,
\eq
where $\Lambda$ denotes the cosmological constant.

The variations of the action with respect to $N, N^i, A, \varphi$ and $g_{ij}$, yield, respectively, the Hamiltonian,
momentum, $A$-,  $\varphi$- constraints, and dynamical equations \footnote{In the expression of $F^{ij}_{a}$ given by
Eq.(3.22) in \cite{ZSWW}, the coefficients $\beta_{s}$ should be replaced by $\hat\beta_{s}$, where $\hat\beta_{s} = (-\beta_0, \beta_n)$
with $n = 1, 2, ..., 8$, while $\hat\gamma_9$ defined in Eq.(3.24)
should be defined as $\hat\gamma_9 = \gamma_5$, instead of $\gamma_5/2$.}, given explicitly  in \cite{ZSWW}.

\section{Spherical Static Vacuum Solutions}
\renewcommand{\theequation}{3.\arabic{equation}} \setcounter{equation}{0}

In this paper, we consider the spherically symmetric static spacetimes,  described by \cite{GPW,GSW},
 \bqn
 \label{4.1}
 && N = N(r),\;\;\; N^i = h(r)\delta^{i}_{r},\;\;\; A = A(r),\nb\\
 &&
 g_{ij} dx^i dx^j = \frac{dr^2}{f(r)}  + r^2d\Omega^2,\;\;\;
 \;\; \varphi = \varphi(r),
 \eqn
where  $d\Omega^2\equiv d\theta^2+\sin^2\theta d\phi^2$.
Using the U(1) gauge freedom, without loss of the generality,  we set 
 \bq
 \lb{4.2}
 \varphi = 0.
 \eq
Then,  we find that ${\cal{L}}_{\varphi}  = 0$.

 In the IR, the spatial curvature is very small, and its high-order derivatives are negligible, so we can safely assume that
 $ {\cal{L}}_{V}$ has only three non-vanishing terms,   given by
 \bq
 \lb{4.3}
 {\cal{L}}_{V}  =  2\Lambda  - R - \beta_{0}a_{i}a^{i}.
 \eq
In the solar system, the effects of cosmological constant are negligible. In addition, the curvature is extremely small.
So, without loss of generality, we can safely set   $\Lambda = \Lambda_g = 0$.
Then, we find that in the present case there are only four-independent equations, which can be taken as the Hamiltonian and momentum constraints,
the A-constraint, obtained from the variation of the gauge field $A$,   and the rr-component of the dynamical
equations, given, respectively, by
  \bqn
 \lb{Hamiltonian}
&& \frac{h^2}{4fN^2}\left[\left(\frac{f'}{f} - \frac{2h'}{h}\right)^2 - \lambda\left(\frac{f'}{f} - \frac{2h'}{h} - \frac{4}{r}\right)^2 + \frac{8}{r^2}\right]\nb\\
&& ~~~ + \beta_0\left[2\frac{N''}{N} - \left(\frac{N'}{N}\right)^2 + \left(\frac{f'}{f} + \frac{4}{r}\right)\frac{N'}{N}\right]    = 0, ~~~~~~ \\
  \lb{Momentum}
&&  \left(\lambda -1\right)\Bigg\{2h'' - \left(\frac{f'}{f} + \frac{2N'}{N} - \frac{4}{r}\right)h' \nb\\
&&~~~~~~~~~~~  - \left[\frac{f''}{f} - \frac{{f'}^2}{f^2} - \frac{f'N'}{fN} + \frac{4}{r}\left(\frac{N'}{N} + \frac{1}{r}\right)\right]h\Bigg\}\nb\\
&& ~~~~~~~~~~~  + \frac{2h}{r}\left(\frac{f'}{f} - \frac{2N'}{N}\right)  = 0,
\\
  \lb{A-Constraint}
&&  (rf)' =1 ,\\
\lb{dynrr}
&&  \left(\frac{A}{f^{1/2}}\right)' = Q(r),
 \eqn
 where
 \bqn
 \lb{4.5a}
 Q(r) &=&
 -\frac{(\lambda-1)rh^2}{16f^{3/2}N}\Bigg[4\left(\frac{f''}{f} - \frac{2h''}{h}\right) \nb\\
 && - \frac{f'}{f}\left(\frac{3f'}{f} + \frac{4N'}{N} + \frac{8}{r}\right)\nb\\
 && + \frac{4h'}{h}\left(\frac{h'}{h} + \frac{2N'}{N}\right) + \frac{16}{r}\left(\frac{N'}{N} + \frac{2}{r}\right)\Bigg]\nb\\
 && +\frac{h^2}{f^{3/2}N}\left(\frac{f'}{f} - \frac{h'}{h} - \frac{N'}{N} - \frac{1}{2r}\right)\nb\\
 && + \frac{N}{2f^{1/2}}\left(\frac{2N'}{N} - \frac{f'}{f}\right)
  + \frac{\beta_0 r}{4f^{1/2}N}{N'}^{2}.
 \eqn
Therefore, in the present case, we have four independent  ordinary differential equations  for four unknowns, $N, \; f, h$ and $A$.
In particular, Eq.(\ref{A-Constraint}) has the general solution,
\bq
\lb{4.5b}
f(r) = 1 - \frac{2B}{r},
\eq
where $B$ is an integration constant.  Once $f$ is known, from Eqs.(\ref{Hamiltonian}) and (\ref{Momentum}) one can find $N$ and $h$, while Eq.(\ref{dynrr}) yields,
\bq \lb{4.6b}
A(r) = \sqrt{1 - \frac{2B}{r}}\left(A_0 + \int{Q(r) dr}\right),
\eq
where $A_0$ is an integration constant. Therefore, the main task now reduces to solve Eqs.(\ref{Hamiltonian}) and (\ref{Momentum})  for $N$ and $h$.
In the following,   let us consider   the two cases $h = 0$ and $h \not= 0$, separately.

 \subsection{$h = 0$}

 In this case,  the momentum constraint is satisfied identically, while Eq.(\ref{Hamiltonian})    reduces   to
 \bqn
 \lb{5.1a}
 \beta_0\left[2\frac{N''}{N}   + \left(\frac{f'}{f} - \frac{N'}{N} + \frac{4}{r}\right)\frac{N'}{N}\right]   = 0.
\eqn
When $\beta_0 = 0$, it is satisfied identically, while Eq.(\ref{4.6b}) yields,
\bq
\lb{5.4}
N = N_{0}\sqrt{1 - \frac{2B}{r}} + A(r),
\eq
where $N_0$ is a constant, and  $A(r)$  is undetermined.
In particular, when $A = 0$, the above solutions reduce to the Schwarzschild  solution,
 \bq
 \lb{5.4a}
 f = N^2 = 1 - \frac{2B}{r},\;\;\; A(r) = 0.
 \eq
Note that in writing the above expression, we had set $N_0 = 1$ by rescaling  $t$. It is remarkable  to note that the above solutions are
independent of the coupling constant $\lambda$. That is, in contrast to all other models of the HL theory \cite{reviews},
the  Schwarzschild  solution (\ref{5.4}) is a solution of the HL theory not only for $\lambda = 1$ but also for any value
 of $\lambda$ in the IR limit. In addition, in the nonprojectable HL theory without the U(1) symmetry, to solve the instability
 problem of the spin-0 gravitons, $\beta_0$ is necessarily non-zero,  $ \beta_0 \not= 0$. Otherwise,
 the spin-0 gravitons become unstable. However, in the current case the U(1) symmetry eliminates the
 spin-0 gravitons \cite{ZSWW}, so the instability problem is out of question, and the physically viable region of the phase
 space includes the point $\beta_0 = 0$. Furthermore, in the projectable case, we have $N = N(t)$, and the solutions with $h = 0$
 take the Schwarzschild form only when the gauge field $A$ and the Newtonian pre-potential $\varphi$ are part of the metric
 \cite{HMT,LMW}. The above observations are  important, when we consider the solar system tests.

When $\beta_0 \not= 0$,   from Eq.(\ref{5.1a}) we find that
\bq
\lb{5.2}
N(r) = \left(N_{0}  + N_1 \sqrt{1 - \frac{2B}{r}}\right)^2,
\eq
where $N_1$ is another  integration constant.  Inserting it into Eqs.(\ref{4.5a}) and (\ref{4.6b}), we obtain
\bqn
\lb{5.3}
A(r)  &=&   N_{0}^2  + A_0\sqrt{1-\frac{2B}{r}}   -N_1^2\frac{B}{r}\nb\\
&& - (\beta_0-1){N}_1^2\left(1-\frac{B}{r}\right).
\eqn

 \subsection{$h \not= 0$}

In this case, to solve Eqs.(\ref{Hamiltonian}) and (\ref{Momentum}), we consider the cases $\lambda = 1 $ and $\lambda \not= 1$, separately.

\subsubsection{$\lambda = 1$}

%

When $\lambda = 1$, Eq.(\ref{Momentum}) has the general solution,
\bq
\lb{5.7}
N = N_0\sqrt{f}.
\eq
Then,   Eq.(\ref{Hamiltonian}) and
Eq.(\ref{4.6b}) yield,
\bqn
 \lb{5.8}
h(r) &=&  N_0 \sqrt{\frac{r-2B}{12r^2}} \Bigg[C_1  + 3B\beta_0\ln\left({\frac{r}{r-2B}}\right)\Bigg]^{1/2},\nb\\
A(r) &=& - \frac{N_0\beta_0}{8r\sqrt{1-\frac{2B}{r}}}\left[2B- (r -2B)\ln\left({\frac{r}{r-2B}}\right)\right]\nb\\
&& + A_0\sqrt{1-\frac{2B}{r}},
\eqn
where $C_1$ is a constant.

Note that, when $B = 0$, the above solutions reduce to
 \bq
 \lb{5.5}
 N = N_0,\;\;\; h=\pm\sqrt{\frac{r_g}{r}}, \;\;\; A = A_0,
 \eq
where $r_g$ is an integration constant. This is precisely    the Schwarzschild solution, but written in the  Painleve-Gullstrand
coordinates \cite{GP}.

\subsubsection{$\lambda \not= 1$}

In this case, let us consider  the two cases, $B = 0$ and $B \not= 0$,  separately.

{\bf Case (i) $\; B=0$:}  Now Eqs. (\ref{Momentum}) and  (\ref{Hamiltonian}) reduce, respectively, to
\bqn
\lb{5.9a}
&& \varepsilon h'' - \varepsilon\left(\frac{N'}{N} - \frac{2}{r}\right)h' - \frac{2}{r}\left[(1+\varepsilon)\frac{N'}{N} + \frac{\varepsilon}{r}\right] h = 0,\nb\\
\\
\lb{5.9b}
&& \frac{h^2}{N^2}\left[\varepsilon r^2 \frac{{h'}^2}{h^2} + 4(1+\varepsilon)r \frac{h'}{h} + 2(1+2\varepsilon)\right] \nb\\
&& ~~~~~ - \beta_0\left(2r^2\frac{N''}{N} - r^2\frac{{N'}^2}{N^2} +
4r\frac{N'}{N}\right)   = 0, \eqn where $\varepsilon\equiv
\lambda-1$.

When $\beta_0=0$, the solutions is given by
\bqn
\lb{5.10a}
 h(r)&=&H_0r^{\alpha_{\pm}},\;\;\;
 N(r)=N_0r^{\beta_{\pm}},\nb\\
A(r)&=&\bar{A}_0 +E_1(r) +\varepsilon E_2(r),
 \eqn
where  $\bar{A}_{0}$ and $H_0$ are constant, and
\bqn
 \lb{5.10aa}
\alpha_{\pm}&=&-2-\frac{2\pm
\sqrt{4+6\varepsilon}}{\varepsilon},\nb\\
\beta_{\pm}&=& \frac{-2+\alpha_{\pm}+\alpha_{\pm}^2}{2+(2+\alpha_{\pm})\varepsilon}\varepsilon\nb\\
 E_1(r)&=&\frac{1}{4N_0r^{\beta_{\pm}}(2\alpha_{\pm}-\beta_{\pm})}\Big[N_0^2r^{2\beta_{\pm}}(2\alpha_{\pm}-\beta_{\pm})\nb\\
 &&\times (4+\beta_{\pm}\beta_0) -2H_0^2r^{2\alpha_{\pm}}(1+2\alpha_{\pm}+2\beta_{\pm})\Big],\nb\\
E_2(r) &=&
H_0^2\frac{(2+\alpha_{\pm})(\alpha_{\pm}-4-2\beta_{\pm})}{4N_0(2\alpha_{\pm}-\beta_{\pm})}r^{2\alpha_{\pm}-\beta_{\pm}}.
\eqn
It is interesting to note that, taking the ``-" sign in the
above solutions, they reduce to the Schwarzschild solution when
$\epsilon \rightarrow 0$. This is in contrast to the case with the
projectability condition $N = N(t)$ \cite{LMW}, in which it was found that such
relativistic limit does not exist.

When $\beta_0\neq0$, it is found  difficult to obtain exact solutions. Instead, we consider the cases where  $\varepsilon$ is very  small,
as one would expect that physical viable solutions must be very close to that of GR where $\lambda_{GR} = 1$. Thus, expanding
$h(r)$, $N(r)$ and $A(r)$ in terms of $\epsilon$, we find that
\bqn \lb{5.11}
h(r)&=&\pm\sqrt{\frac{r_g}{r}}+\varepsilon \left[\frac{H_0}{\sqrt{r}}\mp\frac{9(N_0^2\beta_0r+r_g \ln(r))}{16\sqrt{r_gr}}\right]\nb\\
&& +\mathcal {O}(\varepsilon^2),\nb\\
N(r)&=&N_0+\varepsilon \left[N_1-\frac{9}{8}N_0\ln(r)\right]+\mathcal {O}(\varepsilon^2),\nb\\
A(r)&=&A_0+\varepsilon\frac{9}{16}N_0(\beta_0-2)\ln(r)+\mathcal
{O}(\varepsilon^2).
\eqn
Hence,
\bq
\lb{5.11a}
h(r) \rightarrow \mp  \frac{9}{16}N_0^2\beta_0 \epsilon \sqrt{\frac{r}{r_g}},
\eq
as $r \rightarrow \infty$. Thus, unless $ \beta_0 \epsilon = 0$, the solutions are not asymptotically flat. When $\beta_0 = 0$, it reduces to the
last case, and the corresponding solutions are given by Eq.(\ref{5.10a}). When $\epsilon = 0$, the corresponding solutions are given by
Eqs.(\ref{5.7}) and (\ref{5.8}). Therefore, the case  $\epsilon \beta_0 \not= 0$ has no physically viable solutions, and  in the following we shall not
consider it any further.

{\bf Case (ii) $B \not= 0$:}  In this case, to obtain exact solutions is found also very difficult, and instead we expand
$h(r)$, $N(r)$ and $A(r)$ in terms of $\epsilon$, and find that
\bqn
 \lb{5.12a}
N(r)&=&\hat{N}(r)+\varepsilon \bar{N}(r)+\mathcal {O}(\varepsilon^2),\nb\\
h(r)&=&\hat{h}(r)+\varepsilon \bar{h}(r)+\mathcal {O}(\varepsilon^2),\nb\\
A(r)&=&\hat{A}(r)+\varepsilon\sqrt{f}\int \bar{Q}(r)dr+\mathcal
{O}(\varepsilon^2),
\eqn
where $\hat{h}(r)$, $\hat{N}(r)$ and $\hat{A}(r)$ are the solution
for $\varepsilon=0$,  which are given by Eqs.(\ref{5.7}) and (\ref{5.8}), while
\bqn
\lb{5.12b}
\bar{N}(r)&=&\sqrt{f}\left(N_1+\int N_a(r)dr\right),\nb\\
\bar{h}(r)&=&\frac{f}{r\hat{h}(r)}\left(H_1+\int h_a(r)dr\right),\nb\\
N_a(r)&=&\frac{N_0r}{2\hat{h}}\hat{h}''+N_0\frac{(3B-r)\hat{h}'}{(2B-r)\hat{h}}\nb\\
&& -N_0\frac{5B^2-8Br+2r^2}{2r(r-2B)^2},\nb\\
h_a(r)&=&N_0\beta_0r\sqrt{f}\left(\frac{r}{2}\bar{N}''+\bar{N}'\right)-\frac{r^2}{4f}(\hat{h}')^2\nb\\
&&+\left[\frac{2f-1}{N_0f^{\frac{5}{2}}}\bar{N}-\frac{(1-5f)^2}{16f^3}\right]\hat{h}^2\nb\\
&&+\left[N_0(1-5f)+8\sqrt{f}\bar{N}\right]\frac{r\hat{h}\hat{h}'}{4N_0f^2}\nb\\
&& +\frac{\beta_0N_0B^2\bar{N}}{2r^2f^{\frac{3}{2}}}\nb\\
\bar{Q}(r)&=&\frac{N_0(5-2f-35f^2)+8(\bar{N}-2r\bar{N}')f^{\frac{3}{2}}}{16N_0^2rf^4}\hat{h}^2\nb\\
&&+\frac{1}{N_0f^2}\left(\frac{1-2f}{rf}\hat{h}\bar{h}-\hat{h}\bar{h}'+\frac{r}{2}\hat{h}''\hat{h}\right)\nb\\
&&+\left(\frac{-B}{2N_0rf^3}+\frac{\bar{N}}{N_0^2f^{\frac{5}{2}}}\right)\hat{h}'\hat{h}+\frac{B\bar{N}}{rf^{\frac{3}{2}}}\nb\\
&&+\frac{\bar{N}'}{\sqrt{f}} -\frac{\hat{h}'}{N_0f^2}\left(\bar{h}+\frac{r}{4}\hat{h}'\right)\nb\\
&&+\frac{\beta_0B}{4r^2f^{\frac{5}{2}}}\left(B\bar{N}+2rf\bar{N}'\right).
\eqn
From the above expressions, we can see that the solutions are also not asymptotically flat. In particular, we have
$\bar{N}(r) \propto \ln(r)$. Therefore, this class of solutions is also physically discarded.

 \section{Solar System Tests}
\renewcommand{\theequation}{4.\arabic{equation}} \setcounter{equation}{0}

In the solar system tests, the metric is usually cast in the diagonal form,
\bq
\lb{3.1}
 ds^{2} = - e^{2\Psi(r)}d\tau^2 + e^{2\Phi(r)}dr^2 + r^2d\Omega^2,
\eq
where $\Psi$ and $\Phi$ of the gravitational
 field, produced by a  point-like and motion-less particle with mass $M$,   are parametrized in the forms,
\bqn
\lb{3.2}
 e^{2\Psi}  &=&1-2\chi  +2(\beta-\gamma)\chi^2+...,\nb\\
 e^{2\Phi}  &=&1+2\gamma\chi + ...,
\eqn
where $\beta$ and $\gamma$ are the Eddington parameters.    For the Schwarzschild solution, we have
$\beta_{GR} = \gamma_{GR}  = 1$.

In the solar system,
we have $r_g \equiv GM_{\bigodot}/c^2 \simeq 1.5$ km, and its radius is $r_{\bigodot} \simeq 1.392\times 10^{6}$ km. So, within the solar system
the dimensionless quantity $\chi [\equiv GM/(rc^2)]$   in most cases  is much less than one,  $\chi  \leq r_{g}/r_{\bigodot} \le 10^{-6}$.
The Shapiro delay of the   Cassini probe \cite{BT}, and the solar system ephemerides \cite{WTB} yield, respectively, the bounds \cite{RJ},
\bqn
\lb{3.3}
\gamma - 1&=& (2.1 \pm 2.3)\times 10^{-5},\nb\\
\beta - 1&=& (-4.1  \pm 7.8)\times 10^{-5}.
\eqn

To fit the solutions found in the last section with above observations, let us consider the cases $h = 0$ and $h \not= 0$, separately.

\subsection{$h = 0$}

In this case, let us first consider the solutions given by Eqs.(\ref{4.5b}) and (\ref{5.4}). Expanding $A(r)$ in the form,
\bq
\lb{3.3a}
A(r) = A_0 + A_1\chi + A_2 \chi^2 + {\cal{O}}\left(\chi^3\right),
\eq
we find that 
\bqn
\lb{3.3b}
N^2(r) &=& (N_0+A_0)^2\left[1- 2\frac{N_0\sigma - A_1}{N_0+A_0}\right]\chi\nb\\
 &&+ \big(2N_0A_2 -2\sigma N_0 A_1 + A_1^2+2A_0A_2\nb\\
 && ~~~~ -N_0A_0\sigma^2\big)\chi^2   +  {\cal{O}}\left(\chi^3\right),\nb\\
f^{-1}(r) &=& 1 + 2\sigma\chi + 4\sigma^2\chi^2 +  {\cal{O}}\left(\chi^3\right),
\eqn
where $\sigma \equiv \frac{Bc^2}{GM}$. Note that in writing the above expressions, we can set $t\rightarrow (N_0+A_0)t$.
Comparing  Eq.(\ref{3.3b}) with Eq.(\ref{3.2}), we find that  
\bqn
\lb{3.3bb}
A_1 &=& N_0(\gamma - 1)-A_0 ,\nb\\
 A_{2} &=&  (N_0+A_0)(\beta - \gamma) \nb\\
 && + \frac{N_0}{2}\left(\gamma +1\right)\left(\gamma -1\right)-\frac{A_0}{2} ,\nb\\
\sigma&=&  \gamma . \eqn
Without loss of generality, we can alway set $N_0=1$ and $A_0=0$, for which we find 
\bqn
\lb{3.3c}
A_1 &=& \gamma - 1 \leq 10^{-5},\nb\\
 A_{2} &=&  (\beta - \gamma) + \frac{ \left(\gamma +1\right)}{2}\left(\gamma -1\right) \leq 10^{-5},\nb\\
\sigma&=&  \gamma \simeq 1 + (2.1 \pm 2.3)\times
10^{-5}. \eqn

For the solutions given by Eqs.(\ref{4.5b}) and (\ref{5.2}), we have
\bqn
\lb{3.3d}
N^{2}(r) &=& \left(N_0 + N_1\right)^4 - 4\sigma N_1\left(N_0 + N_1\right)^3\chi \nb\\
&& + 2 \sigma^2N_1\left(2N_1 -N_0\right)\left(N_0 + N_1\right)^2\chi^2 \nb\\
&&+  {\cal{O}}\left(\chi^3\right),
\eqn
where $f(r)$ is still given by Eq.(\ref{3.3b}). Then, comparing them with Eq.(\ref{3.2}), we obtain
\bqn
\lb{3.3e}
&& 2\beta - \gamma = \frac{3}{2},\;\; \gamma =  \sigma,\nb\\
&& N_0 = \frac{2\gamma - 1}{2\gamma},\;\;\; N_1 = \frac{1}{2\gamma}.
\eqn
Clearly, this class of solutions does not satisfy the solar system tests,   and must be discarded.

\subsection{$ h \not= 0$}

To study the solar system tests for this class of solutions,   we need first transform the   experimental results of Eqs.(\ref{3.2}) and (\ref{3.3})
into the form,
\bq
\lb{3.5}
ds^2=-N^2(r)
dt^2+\frac{1}{f(r)}\Big(dr+h(r)dt\Big)^2+r^2d\Omega^2,
\eq
which can be done by    the coordinate transformations,
\bq
\lb{3.4}
\tau=t-\int^r e^{-\Psi}\sqrt{e^{2\Phi}-f^{-1}}dr,
\eq
where $f$ is given by Eq.(\ref{4.5b}), and
\bq
\lb{3.6}
N = e^{\Psi+\Phi}\sqrt{f},  \;\;\;
h = e^{\Psi}\sqrt{f^2e^{2\Phi}-f},
\eq
or inversely,
\bq
\lb{3.6a}
e^{2\Psi} =  N^2 - \frac{h^2}{f},\;\;
e^{2\Phi} = \left(N^2 - \frac{h^2}{f}\right)^{-1}\frac{N^2}{f}.
\eq
It should be noted that, although the coordinate transformations (\ref{3.4}) are forbidden by the foliation-preserving  diffeomorphisms, we
assume that experimental results can be expressed freely in any coordinate systems.

Now let us first consider the solutions given by
%
Eqs.(\ref{5.7}) and (\ref{5.8}), from which we find that
\bqn
\lb{3.8}
e^{2\Psi} &=&   1-\left(2+\frac{C_1}{12B}\right)\sigma\chi-\beta_0\frac{\sigma^2}{2}\chi^2+\mathcal {O}(\chi^3),\nb\\
e^{2\Phi} &=&  1+\left(2+\frac{C_1}{12B}\right)\sigma\chi+\mathcal{O}(\chi^2).
\eqn
Note that in writing the above expressions, we set $N_0 = 1$.
Comparing the above expressions with Eq.(\ref{3.2}), we find that
\bqn
\lb{3.9}
&& \gamma = 1, \;\;\; C_1 =  24\left(\frac{GM}{c^2} - B\right), \nb\\
&& \beta -  1  = - \left(\frac{c^2}{2GM}\right)^2\beta_0 B^2.
\eqn
Clearly, by properly choosing $\beta_0$ and $B$, the conditions (\ref{3.3}) can be easily satisfied  for this class of solutions.

For the solutions (\ref{5.10a}), we find that
\bqn
\lb{3.9a}
e^{2\Psi} &=&   
N_0^2\left(\frac{B}{\sigma\chi}\right)^{2\beta_{\pm}}-H_0^2\left(\frac{B}{\sigma\chi}\right)^{2\alpha_{\pm}}\Bigg[1+2\sigma\chi\nb\\
&&+4\sigma^2\chi^2+\mathcal {O}(\chi^3)\Bigg],\nb\\
e^{2\Phi}
&=&N_0^2\left(\frac{B}{\sigma\chi}\right)^{2\beta_{\pm}}\Bigg[N_0^2\left(\frac{B}{\sigma\chi}\right)^{2\beta_{\pm}}\left(1-2\sigma\chi\right)\nb\\
&&-H_0^2\left(\frac{B}{\sigma\chi}\right)^{2\alpha_{\pm}}\Bigg]^{-1}.
\eqn
Comparing the above with Eq.(\ref{3.2}), we find that
\bqn
\lb{3.9b}
 \alpha_{\pm}=-\frac{1}{2}, \;\;\; \beta_{\pm}=0.
\eqn
 This means $\lambda=1$. Thus, in this case the solutions are consistent with   the solar system tests only when $\lambda = 1$.

 \section{Conclusions}
\renewcommand{\theequation}{5.\arabic{equation}} \setcounter{equation}{0}

In this paper, we have studied the observational constraints  of  the nonprojectable HL theory with the enlarged symmetry (\ref{symmetry})
within the solar system. In particular,  we have first found the static spherical vacuum solutions of the theory in the IR limit, and then considered
their post-Newtonian limits. Because of the breaking of the Lorentz symmetry of the theory, the Birkhoff theorem does not hold here,
and there exist several families of vacuum solutions that are asymptotically flat. Among them, only two families of these solutions
satisfy the observational constraints  imposed on the Eddington-Robertson-Schiff parameters, $\gamma$ and $\beta$, 
by properly choosing the free parameters of the solutions. One is given by Eqs.(\ref{4.5b}), (\ref{5.4}) with an arbitrary function $A(r)$,
subjected to the conditions (\ref{3.3c}).  This class of solutions is diagonal and valid for any $\lambda$.
The other is non-diagonal ($h \not= 0$), given by Eqs.(\ref{4.5b}), (\ref{5.7}) and (\ref{5.8}) with  $\lambda = 1$, and subjected to  the constraints (\ref{3.9}).

It should be noted that,  in contrast to the projectable HL theory with the enlarged symmetry (\ref{symmetry}) \cite{HMT,WWa,daSilva,HW}, the
consistency is achieved
without the gauge field  and Newtonian prepotential being  part of the metric  \cite{LMW,HMT}.

A remarkable feature is that the Schwarzschild solution (\ref{5.4a}) written in the Schwarzschild coordinates is also a physically viable solution of the
nonprojectable HL theory with the enlarged symmetry (\ref{symmetry}) in the IR not only for $\lambda = 1$ but also for any value of $\lambda$.
This is different from all other models of the HL theory proposed so far \cite{reviews}.

Applying the general results presented in \cite{Wang} to  the physically viable spherical vacuum solutions obtained in this paper, one can easily
construct slowly rotating vacuum solutions of  the nonprojectable HL theory with the enlarged symmetry (\ref{symmetry}).  Some of such constructed
solutions clearly represent slowly rotating black holes   in the IR limit \cite{GLLSW}.

Post-Newtonian approximations are generally characterized by ten parameters \cite{Will}. The studies of observational constraints in the nonprojectable
HL theory without the local U(1) symmetry \cite{BPS} showed that the preferred frame effects  imposed the most stringent constraints on
the suppression energy $M_{*}$. The structures of the nonprojectable HL theory with and without the local U(1) symmetry are quite different. In particular,
the spin-0 gravitons do not exist in the current model. Thus, it would
be very interesting to see which kind of constraints the preferred frame effects can impose on $M_{*}$ as well as on  other coupling constants of the
theory. We wish to come back to this issue soon in another occasion.

\section*{\bf Acknowledgements}

This work was supported in part by DOE  Grant, DE-FG02-10ER41692 (AW),  NSFC No. 11173021 (AW), NSFC No.11075141 (AW),
FAPESP No. 2012/08934-0 (KL); NSFC No. 11178018 (KL); and NSFC No. 11075224 (KL).


\end{document}